\documentclass{moriond_my}

\usepackage{amsmath,amssymb}
\usepackage[utf8]{inputenc}
\usepackage{cite}

\newcommand{\Photo}{}

\begin{document}
\vspace*{4cm}
\title{From Swampland to Phenomenology and Back\footnote{Contribution to the proceedings
for the 2019 EW session of the 54th ``Rencontres de Moriond''. The slides for the talk are available from 
\url{http://member.ipmu.jp/masahito.yamazaki/files/2019/Moriond_2019.pdf}. In the original version of this note I stated ``Since this proceeding is limited to strictly $6$ pages with references included, we will not have space to list all the relevant references." In the revised version some more references have been added, but I would still encourage interested readers to refer to the INSPIRE-HEP database for a complete list.}}

\author{Masahito Yamazaki}

\address{Kavli Institute for the Physics and Mathematics of the Universe (WPI), \\
University of Tokyo, Kashiwa, Chiba 277-8583, Japan}

\maketitle

\abstracts{Swampland conjectures are a set of proposed necessary conditions for a low-energy effective field theory to have a UV completion inside a theory of quantum gravity. Swampland conjectures have interesting phenomenological consequences, and conversely phenomenological considerations are useful guidelines in sharping our understanding of quantum gravity.}

\section{Swampland Conjectures}

During the Moriond electroweak session we have seen many examples of fruitful interplays between theory and experiment---theorists try to explain experimental findings, while experimentalists turn to theorists for useful guidelines for what to look for. 

From a theorist's viewpoint, coming up a ``theory'' often involves picking up a certain low-energy quantum field theory (QFT).
One can then discuss experimental constraints on the parameters of the theory, possible new signatures predicted by the theory, and so on.
The big problem, however, is that there are tremendous numbers of possible quantum field theories you can write down, and hence one has to face with huge numbers of possibilities.

\begin{figure}[htbp]
\centering\includegraphics[scale=0.45]{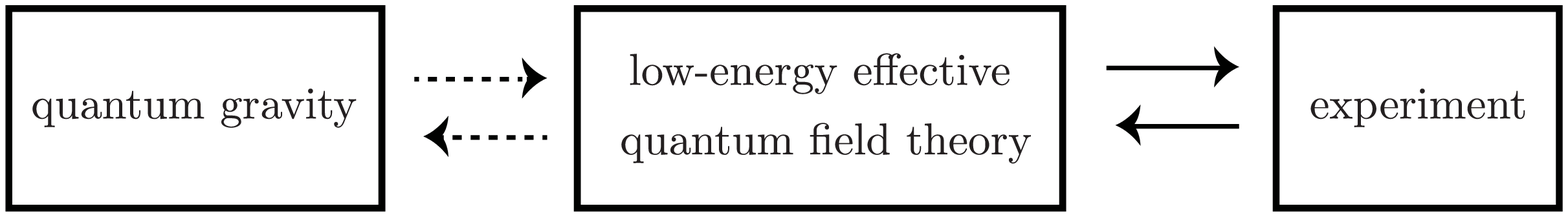}
\caption{A phenomenologist often comes up with a specific low-energy QFT and works out its experimental consequences. It is often not clear, however, which QFT one should start with, since there are simply too many possibilities. 
If one starts with quantum gravity, one sometimes finds good evidence that certain low-energy QFTs are \textrm{not possible}.
This is the content of the swampland conjectures. Such conjectures help to narrow down possibilities, and could reduce the the difficult job of a ``low-energy'' phenomenologist.}
\label{topdown}
\end{figure}

One might therefore wish that there are new ingredients in theorist's toolkit.
I am going to argue that such a new tool comes from quantum gravity, albeit in somewhat unconventional forms (Fig.~\ref{topdown}). 

Quantum gravity in itself is typically believed to be located at an
extremely high energy scale. We can, however, ask whether or not a given low-energy effective field theories can be reproduced from a theory of quantum gravity.

Traditionally, this involves a specific string theory setup---this can be heterotic string theory on Calabi-Yau three-fold, F-theory on elliptically-fibered Calabi-Yau four-fold, M-theory on $G_2$ holonomy manifolds, etc.
By contrast we are asking a more {\it universal} question, at least in spirit: does the low-energy effective field theory
have a UV completion in quantum gravity? If the answer is yes, we say that the theory is in the \textrm{landscape}, while if no 
we say that the theory is in the \textrm{swampland}.\cite{Vafa:2005ui,Ooguri:2006in} In this terminology (by C.~Vafa), 
if you are a ``low-energy'' phenomenologist you need to make sure that your favorite theory is in the landscape, and not in the swampland.

\begin{figure}[htbp]
\centering\includegraphics[scale=0.45]{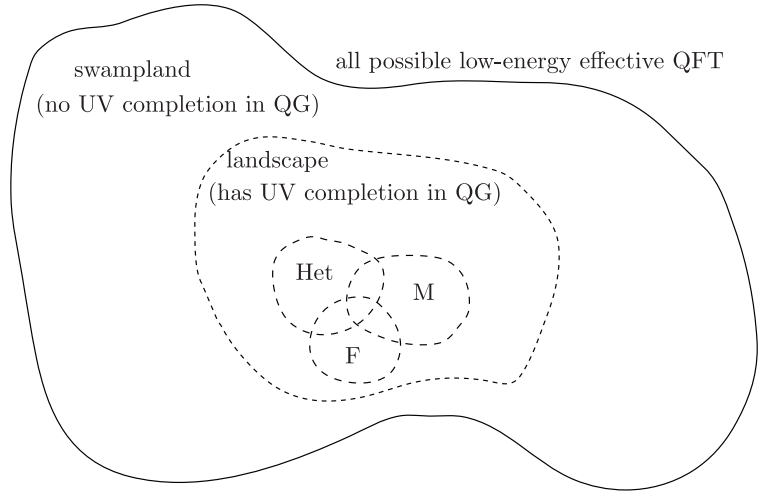}
\caption{Inside the space of possible low-energy QFTs, some are realized by specific string theory constructions,
such as Calabi-Yau three-fold compactifications of the Heterotic theory. The consequences from these specific constructions,
may well be setup-dependent, and might not be obeyed in other string theory constructions. In the swampland conjectures,
by contrast, one tries to formulate \textit{universal} consequences from the existence of UV completion with gravity.
If such conjectures are true, that has some interesting implications to low-energy model building.
There are indications that the constraints from swampland conjectures are so strong that 
``landscape'' is actually tiny inside the space of all possible low-energy QFTs.}
\end{figure}

One might ask ``isn't quantum gravity notoriously difficult''? For proper treatment of quantum gravity,
you have to wait for your quantum gravity colleagues to solve it, and as of this writing there is no indication
that this will happen in anytime soon. Does this mean that you can forget about this altogether?

Despite the lack of full understanding of quantum gravity, we do have some useful clues about quantum gravity.
First, we know about semiclassical description of black holes, such as that black holes have Bekenstein-Hawking entropy,
that black holes evaporate with Hawking radiation.\cite{Hawking:1974sw} 
We can devise many gedanken-experiments on 
black holes, and think about the consequences, loopholes, and so on.
Second, there are many examples/data from string theory,\footnote{String theory is a rather promising theory of quantum gravity, and makes it possible to do many quantitative computations, thus providing invaluable data for swampland constraints. For this reason we often use string theory in swampland discussions. One should note, however, that the arguments from semiclassical black holes, as mentioned previously, are independent of string theory. Consistency between black-hole-inspired conjectures and string-theory-inspired conjectures have also been discussed.} e.g.\ results based on 
a specific class of compactification manifolds.

The way to proceed is look at various data available, and formulate \emph{swampland conjectures},
a set of necessary conditions for the existence of UV completion. Once we formulate these conjectures,
we can set these conjectures by looking many more examples, this often leads to quantitative tests.
We can also try to verify the mutual consistency between different conjectures.

There are many swampland conjectures in the market, indeed too many to be listed here and readers are referred to comprehensive reviews.\cite{Brennan:2017rbf,Palti:2019pca} Here, to illustrate the idea let me mention one particular conjecture by M.~Reece \cite{Reece:2018zvv} (see Hebecker et al.\cite{Hebecker:2017uix} for a closely related work). 

Consider a $U(1)$ gauge field $A$, namely a photon, with 
Su\"uckelberg mass $m$.\footnote{This is a gauge-invariant mass of the form $\mathcal{L}\supset m^2 (A_{\mu}- \partial_{\mu} \theta)^2$ for a scalar field $\theta$, with gauge transformation $\delta A_{\mu}=\partial_{\mu} \chi, \delta\theta=\chi$.}
Let us denote the gauge coupling constant by $e$.
Then the conjecture states that  the UV-cutoff $\Lambda_{\rm UV}$ of the theory
should satisfy the inequality
\begin{equation}
\Lambda_{\rm UV} \lesssim \textrm{min} \left( (m M_{\rm Pl}/e)^{\frac{1}{2}},\,  e^{\frac{1}{3}} M_{\rm Pl} \right) \;.
\label{Lambda.conjecture}
\end{equation}
This places interesting constraints on light dark photons,\cite{Reece:2018zvv}; note scenarios involving such light dark photons were discussed during the Moriond conference.\footnote{See, however, Craig et al.\cite{Craig:2018yld} for discussion on possible loopholes.}

This conjecture is motivated by many other swampland conjectures, in particular the distance conjecture\cite{Ooguri:2006in,Klaewer:2016kiy}
and the weak gravity conjecture\cite{ArkaniHamed:2006dz}, where the latter
is an upgrade of the old conjecture that there is no global symmetry in quantum gravity.\cite{Misner:1957mt,Polchinski:2003bq,Banks:2010zn,Harlow:2018tng}
In this respect, the conjecture \eqref{Lambda.conjecture} is an excellent example for an outcome
from interrelations between several different approaches to the swampland program.

\section{De Sitter Swampland Conjecture}

While the swampland program by now has history of more than $10$ years, 
recently there has been renewal of interest in this topic.  One of the reasons for this 
is the paper\cite{Obied:2018sgi}  on the ``de Sitter swampland conjecture'' (hereafter dS conjecture)  in June 2018. This paper immediately generated interest and  and controversies in the community.

The dS conjecture states that the total scalar potential $V$ of any effective field theory\footnote{We choose canonical normalization for scalar fields. We assume there are only finitely many scalar fields in the Lagrangian.}
satisfies an inequality
\begin{align}
M_{\rm Pl}\,  | \nabla V | \ge c \,V \;.
\label{dS}
\end{align}
Here $M_{\rm Pl}\simeq 2\times 10^{18} \textrm{GeV}$ is the Planck scale,
$c$ is an $O(1)$ positive constant.

After the proposal of the conjecture, problems of this conjecture has been pointed out---the dS conjecture excludes local maximum with positive energy
\begin{align}
\nabla V=0 \;, \quad \nabla^2 V<0 \;, \quad V>0 \;.
\end{align}
Such local maximum, however, clearly exists in the potential of the Higgs field,
and of the axion\cite{Peccei:1977ur,Weinberg:1977ma,Wilczek:1977pj} 
(if an axion exists).

This in itself does not exclude the dS conjecture, since one can think of various loopholes---for example, we can 
enlarge the field space by coupling the Higgs/axion to the quintessence field.
However, after several independent analysis (involving many fun ingredients,
such as no-go theorem for electroweak modification, fifth-force constraints, etc.), one comes to the conclusion that most such loopholes can be eliminated, except for some rather exotic scenarios.\cite{Denef:2018etk,Murayama:2018lie,Choi:2018rze,Hamaguchi:2018vtv}

In view of these results, a number of refinements of the dS conjecture have been proposed.\cite{Andriot:2018wzk,Garg:2018reu,Murayama:2018lie,Ooguri:2018wrx,Garg:2018zdg,Andriot:2018mav}
For example, one possible modification\cite{Garg:2018zdg,Ooguri:2018wrx} states that  
\begin{align}
M_{\rm Pl} | \nabla V | \ge c \, V \quad \textrm{or } \quad \textrm{min}(\nabla^2 V) \le -c' \, V \;,
\label{RdS}
\end{align}
where $c$ and $c'$ are $O(1)$ positive constants. This is slightly stronger than the another conjecture,\cite{Murayama:2018lie}
which corresponds to the special value $c'=0$ of the conjecture above. We here call these conjectures
the refined de Sitter conjectures (hereafter RdS conjectures).

While the RdS conjectures evades the problems with the local maxima of the Higgs or axion potential,
they still have an important consequence---the conjectures forbid the stable de Sitter vacua:
\begin{align}
\nabla V=0 \quad \;, \quad \nabla^2 V > 0 \;,  \quad V>0 \;.
\end{align}

This conjecture is therefore tension with the literature (e.g.\ KKLT scenario\cite{Kachru:2003aw}) for de Sitter vacua in string theory (see also recent articles\cite{Sethi:2017phn,Akrami:2018ylq,Kachru:2018aqn}). It seems fair to say that the RdS conjectures are speculative, at least as general statements, and might hold only in the asymptotic region of the moduli space where we have parametric control.\footnote{It would be interesting to fully justify the RdS conjectures even in such weakly-coupled regions of the parameter space (there has recently been attempts in this direction\cite{Ooguri:2018wrx,Hebecker:2018vxz}; note that the classic argument by Dine and Seiberg\cite{Dine:1985he} in itself
does not automatically guarantee this). As we emphasized before, the point of the swampland conjectures
is to work out the consequences which holds irrespective of specific string theory setups, and even the ``weakly-coupled RdS conjecture'' is universal in this sense.} In the following we will assume the RdS conjectures and 
work out the consequences.
This will lead us to interesting scenarios,
and in my opinion it is of value to explore these possibilities, irrespective of the validity of the RdS conjectures---ultimately the finally verdict is up to the Nature.

\subsection{Dark Energy}

If RdS conjectures are correct, dark energy cannot be the cosmological constant.
One possibility is then to consider a dynamical scalar field, the quintessence.\cite{Ratra:1987rm,Wetterich:1987fm,Zlatev:1998tr}

There are enormous challenges in quintessence model building. 
First of all, the quintessence potential should be
extremely flat, to avoid rapid change of the size of the dark energy.
Second, the potential in itself should have a correct size to 
explain the present-day energy scale of dark energy $\Lambda^4 \simeq  10^{-120} M_{\rm Pl}^4  \ll M_{\rm Pl}^4$.

In our recent paper\cite{Ibe:2018ffn}, we looked into the possibility of the  
electroweak quintessence axion, where the quintessence is 
the axion field\cite{Fukugita:1994hq,Frieman:1995pm,Choi:1999xn} 
for the electroweak $SU(2)$ gauge group.\cite{Fukugita:1994hq,Nomura:2000yk,McLerran:2012mm}\footnote{See e.g.\ \cite{Agrawal:2018own,Heisenberg:2018yae,Cicoli:2018kdo,Akrami:2018ylq,Heisenberg:2018rdu,Marsh:2018kub,DAmico:2018mnx,Han:2018yrk,Brandenberger:2018xnf,Yang:2018xah,Olguin-Tejo:2018pfq,Agrawal:2018rcg,Heckman:2018mxl,Chiang:2018lqx,Thompson:2018ifr,Elizalde:2018dvw,Ibe:2018ffn,Tosone:2018qei,Emelin:2018igk,Acharya:2018deu,Hertzberg:2018suv,Raveri:2018ddi,Heckman:2019dsj,Heisenberg:2019qxz,Brahma:2019kch,Kaloper:2019lpl}
for other recent discussion of quintessence and swampland conjectures.} 

If such an electroweak axion exists, it can explain the flatness of the potential
since the shift symmetry of the axion is broken only by non-perturbative instanton effects.
Moreover, the energy scale of the axion potential is given by the dynamical scale, 
which is estimated to be\cite{Nomura:2000yk,McLerran:2012mm}
\begin{align}
\Lambda^4&=M_{\rm Pl}^4 \, e^{-\frac{2\pi}{\alpha_2(M_{\rm Pl})}} \simeq 10^{-130} M_{\rm Pl}^4  \;,
\label{Lambda_a}
\end{align}
where we use the value of the electroweak coupling constant $\alpha_2=g_2^2/(4 \pi)$ at the Planck scale.
This is very close to the present-day scale of the dark energy. Since the axions are often generated by string theory, one hope
such electroweak quintessence axions arise from string theory, and further motivations for the existence of such axion are provided by the RdS conjectures.\cite{Ibe:2018ffn,Dvali:2018dce}

Of course, irrespective of the validity of the RdS conjectures,
the existence of the quintessence can be probed by observation,
from the measurement of the equation-of-state parameter $w=p/\rho$ 
for various values of redshift. While the current results show no sign of deviation from 
$w=-1$,\cite{Aghanim:2018eyx} it is important to continue observational searches in higher precision.

\subsection{Inflation}

What are the consequences of the RdS conjecture for the early universe, namely inflation?

If we insist on the two $O(1)$ parameters $c, c'$ in Eq.~\eqref{RdS} to be $c,c'\sim 1$
(and not $c,c'\sim 0.01$, for example),
then we found\cite{Fukuda:2018haz} that the simplest type of inflationary models,
namely single-field inflationary models with canonical kinetic terms,
can generate inflation with sufficient number of e-foldings, but 
have trouble reproducing the observed value\cite{Akrami:2018odb} of the spectral index.\footnote{Note that 
this conclusion is different from earlier papers (see e.g.\  \cite{Achucarro:2018vey,Kehagias:2018uem,Matsui:2018bsy,Garg:2018reu,Ben-Dayan:2018mhe,Kinney:2018nny,Brahma:2018hrd}) which discussed the consequences of the condition \eqref{dS} (as opposed to the refined version \eqref{RdS})---under \eqref{dS}
single-field inflationary models with canonical kinetic terms have trouble generating a sufficient number of e-foldings.}
The problem can be evaded in multi-field inflationary models,
e.g.\ by the curvaton scenario.\cite{Linde:1996gt,Enqvist:2001zp,Lyth:2001nq,Moroi:2001ct}
This could lead to interesting signatures, such as primordial non-Gaussianities,\cite{Kawasaki:2011pd}
which can be tested in future improved observations of the 
cosmic microwave background radiation.

\section{Summary}

In this paper 
we have introduced the swampland program, which is an attempt
to extract {\it universal} consequences of quantum gravity
in low-energy effective field theory. The swampland conjectures, if correct, have
many implications to physics in vastly different energy scales,
ranging from the energy scale of dark energy all the way up to the 
energy scale of inflation. Conversely, we have presented examples where
the bottom-up phenomenological constraints
affect the on-going discussion about 
swampland conjectures.

\begin{figure}[htbp]
\centering\includegraphics[scale=0.5]{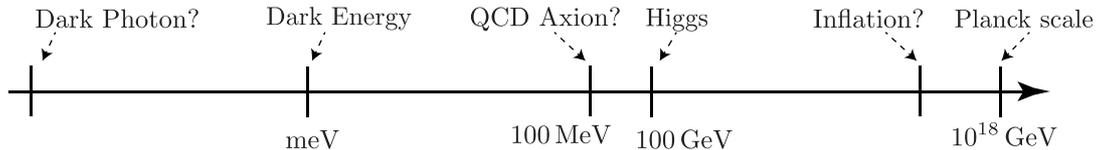}
\caption{While quantum gravity in itself might be associated with very high energy scale (such as the Planck scale),
swampland conjectures originating from quantum gravity have implications phenomena in vastly different energy scales,
as explained in this paper. }
\end{figure}

In this sense, the search for quantum gravity 
is tied with the study of ``low-energy physics'', either in theory and experiment (as expressed in Fig.~\ref{topdown}).
In order to ``make sure that no stones are left unturned''\footnote{Comment from
the theory summary talk by Gudrun Hiller in Moriond EW 2019.}, 
swampland program is a useful approach to keep in mind
for any particle phenomenologist and cosmologist.

\section*{Acknowledgments}

MY would like to thank the organizers of the 54th Rencontres de Moriond
for invitation and for providing stimulating atmosphere. He is partially supported by WPI program (MEXT, Japan) 
and by JSPS KAKENHI Grant No.\ 17KK0087, No.\ 19K03820 and No.\ 19H00689.

\bibliographystyle{nb}
\bibliography{swampland_Moriond_updated}

\end{document}